\documentclass[aps,prx,twocolumn,groupedaddress]{revtex4-1}
\usepackage{amsmath}
\usepackage{graphicx}
\usepackage{color}

\begin{document}

\title{Theory of photoexcited and thermionic emission across a two-dimensional graphene-semiconductor Schottky junction}

\author{Maxim Trushin}
\affiliation{Centre for Advanced 2D Materials, National University of Singapore, 6 Science Drive 2, Singapore 117546}

\date{\today}

\begin{abstract}
This paper is devoted to photocarrier transport across a two-dimensional graphene-semiconductor Schottky junction. 
We study linear response to monochromatic light with excitation energy well below the semiconductor band gap.
The operation mechanism relies on both photoelectric and thermionic emission from graphene to a two-dimensional semiconductor
under continuous illumination and zero bias.
Due to the thermalization bottleneck for low-energy carriers in graphene,
the photoelectric contribution is found to dominate the photoresponse at near-infrared excitation frequencies and below.
The extended thermalization time provides an interesting opportunity to facilitate 
the interlayer photocarrier transport bypassing the thermalization stage.
As a result, the total photoresponsivity rapidly increases with excitation wavelength making
graphene-semiconductor junctions attractive for photodetection at the telecommunication frequency.
\end{abstract}


\maketitle

\section{Motivation}

Following the discovery of graphene \cite{Nature2007geim}, two-dimensional (2D) materials and their junctions
(also known as van der Waals heterostructures \cite{geim2013van}) have experienced a boom over the last decade\cite{Science2016novo}
offering innovative opportunities not possible within conventional semiconductor-based optoelectronics \cite{Nanoscale2015roadmap}.
The central phenomenon employed in optoelectronic devices (photodetectors, photovoltaic cells, etc.) is the conversion of light energy into electricity,
when photons are consumed to deliver photoexcited electrons to an external circuit where they do electrical work.
In a conventional optoelectronic device, the carriers are extracted after they thermalize and dissipate some energy \cite{Nelson2004}.
The optically active van der Waals heterostructures offer an unprecedented opportunity to increase the amount of work done per photon:
Their thickness is so small that the photocarriers can be extracted while they are still hot or even out of the thermal equilibrium \cite{NatPhys2016ma}.
The excess energy could be then utilized for increasing the photoresponsivity,
as it has been recently predicted \cite{NanoLett2016rodriguez} and observed in 
graphene-MoS$_2$ \cite{SciRep2014zhang,ACSNano2016defazio}, graphene-WSe$_2$ \cite{massicotte2016photo},
graphene-WS$_2$ \cite{yuan2018photocarrier} 2D junctions, as well as in
graphene-Si Schottky contacts \cite{NanoLett2016goykhman}.

\begin{figure}
\includegraphics[width=\columnwidth]{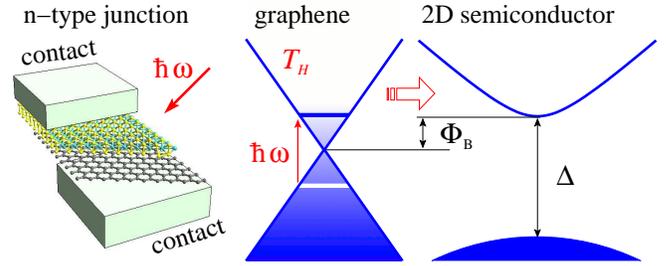}
\caption{\label{fig1} A two-dimensional $n$-type graphene-semiconductor Schottky junction with the potential barrier $\Phi_B$
is continuously illuminated by monochromatic light of frequency $\omega$ with excitation energy $\hbar\omega$ below the semiconductor band gap $\Delta$.
The interlayer photocarrier transport at  $\hbar\omega/2 > \Phi_B$ is due to photoelectric and thermionic emission
provided  correspondingly by the photoexcited (at energy $\hbar\omega/2$) and thermalized (at temperature $T_H$)
steady-state photoelectron distributions in intrinsic graphene. Photoelectric emission turns out to be the dominant photoresponse mechanism
in the near-infrared excitation region due to the extended photocarrier thermalization time, see Fig. \ref{fig2}.}
\end{figure}


\section{Problems solved}

\begin{figure}
\includegraphics[width=\columnwidth]{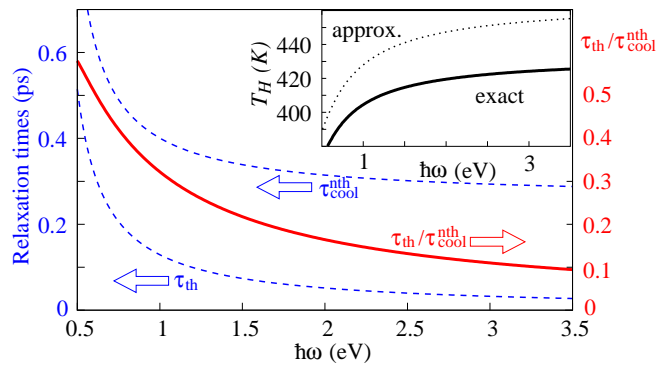}
\caption{\label{fig2} 
Thermalization and energy dissipation times (blue dashed lines) and their ratio (red solid line) for nonthermalized photocarriers in graphene.
At  the excitation energy within the visible spectrum, the two times differ from each other by an order of magnitude.
Within the near-infrared region, on the other hand, the difference is less than a factor of $2$, i. e. thermalization and cooling become two competing processes.
Neither $\tau_\mathrm{th}$ nor $\tau_\mathrm{cool}^\mathrm{nth}$ depends on the absorbed intensity within our linear response model.
The inset shows the corresponding hot carrier temperature within the same excitation energy interval as in the main panel.
The absorbed intensity is constant ($I_a=10$ W/mm$^2$), but the temperature increases
with $\hbar\omega$ because thermalization becomes much faster than energy dissipation.
The exact solution comes from Eq. (\ref{bal-start}).  The approximate solution is given by Eq.~(\ref{lowTH}).}
\end{figure}

\begin{figure}
\includegraphics[width=\columnwidth]{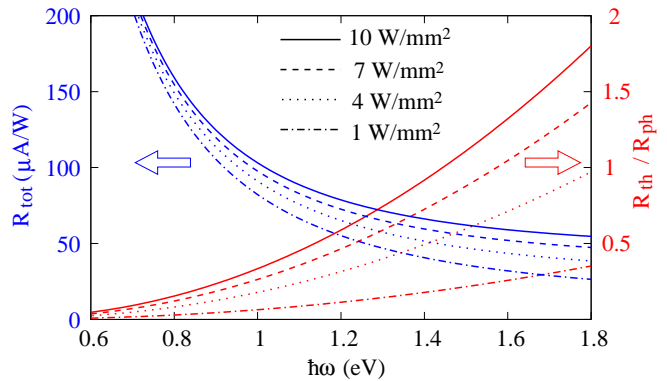}
\caption{\label{fig3} Total zero-voltage photoresponsivity $R_\mathrm{tot}=R_\mathrm{ph} + R_\mathrm{th}$ (blue) given by Eq. (\ref{photores})
and the ratio $R_\mathrm{th}/R_\mathrm{ph}$ (red) given by Eq. (\ref{Ratio}) vs. excitation energy for different absorbed intensities $I_a$.
The Schottky barrier height is about $0.3$ eV; other parameter values are discussed in the main text.
The figure demonstrates that the responsivity in the near-infrared excitation region is mostly due to the non-thermalized photocarriers ($R_\mathrm{ph}$)
and is substantially higher than in the optical regime where thermionic emission ($R_\mathrm{th}$) dominates the photoresponse.}
\end{figure}

We focus on a single junction between intrinsic graphene and a direct-gap 2D semiconductor (see Fig. \ref{fig1}).
Once the junction is continuously illuminated by light with excitation energy below the semiconductor band gap an excess carrier occupation occurs on graphene's side
and, as a consequence, a photocurrent flows across the interface.
We address two theoretical problems in this paper.
(i) The conceptual problem is to identify the ultimate mechanism causing electron transport across a perfectly 2D junction at zero bias voltage.
In the 2D limit, the electrons always remain in their respective layers as they by definition do not have an out-of-surface momentum \cite{liang2015electron,misra2017thermionic}
that is a crucial ingredient in the standard theory of thermionic emission for bulk semiconductors \cite{crowell1965richardson}.
The conventional built-in electric field model seems to be invalid as well \cite{SciRep2014zhang} thanks to the vanishing 
depletion region in perfectly 2D junctions.
In attempt to solve the problem, an external phenomenological parameter is usually introduced
(e.g., an escape time \cite{ryzhii2017nonlinear} $\tau_\mathrm{esc}$,
a transit time \cite{roy2013graphene} $\tau_\mathrm{transit}$, or a tunneling parameter \cite{NanoLett2016rodriguez}).
Here, we assume a perfectly 2D junction with some short-range interface disorder (e.g., remaining organic molecules \cite{PRL2010wehling})
that results in momentum randomization and interlayer hopping. 
The major merit of this model is an obvious correspondence between the in-plane carrier mobility and out-of-plane hopping rate that makes the model self-consistent.
(ii) The technical problem is to derive a steady-state photocarrier distribution function in graphene depending on electron-phonon (e-ph) 
and electron-electron (e-e) scattering parameters as well as on illumination wavelength and intensity. 
A few earlier models \cite{PRL2009bistritzer,PRB2009kubakaddi,PRB2009tse,PRB2012low,PRL2012song,JPCM2015song} describing electron cooling in graphene 
have assumed that the carriers are already thermalized and phonon emission just diminishes carrier temperature. 
The photocarrier dynamics subject to e-e and e-ph scattering has been theoretically investigated in Refs.
\cite{Nanolett2010winzer,PRB2011malic,PRB2011kim,APL2012malic,NJP2013sun,APL2016iglesias}
but no explicit expression for a nonthermalized photocarrier distribution function has been provided until now.
Here, we derive explicit expressions for the thermalization and energy dissipation (cooling) times, which characterize the steady-state photoexcited occupation.
We find that thermalization processes are not always {\em much} faster than energy dissipation,
in marked contrast to conventional bulk metals and semimetals.
As long as excitation energy is within the optical region,
the resulting steady-state occupation turns out to be rather thermalized and is described by the Fermi-Dirac distribution with elevated temperature.
At lower excitation frequencies, the steady-state photocarrier distribution is rather non-thermalized and determined by the spectral shape
of incident radiation.
The physical quantity that reflects this crossover is the internal photoresponsivity determined as a ratio between the photocurrent produced
and radiation power absorbed. We find that the total photoresponsivity, $R_\mathrm{tot} = R_\mathrm{th} + R_\mathrm{ph}$, can be written as a sum of the thermionic
($R_\mathrm{th}$) and photoelectric ($R_\mathrm{ph}$) contributions with its ratio given by 
\begin{equation}
\label{Ratio}
 \frac{R_\mathrm{th}}{R_\mathrm{ph}} = \frac{2}{\pi\hbar} \frac{\omega \Phi_B T_H}{v^2 \tau_\mathrm{th}I_a} \mathrm{e}^{-\frac{\Phi_B}{T_H}},
\end{equation}
where $\hbar$ is the Planck constant, $v\simeq 10^6$ m/s is the carrier velocity in graphene, 
$\Phi_B$ is the Schottky barrier, $\omega$ is the photon frequency ($\hbar\omega/2>\Phi_B$, see Fig. \ref{fig1}), 
$I_a$ is the radiation intensity absorbed, 
$\tau_\mathrm{th}$ is the photocarrier thermalization time due to the e-e and e-ph collisions, i.e.,
$\tau_\mathrm{th}^{-1} = \tau_\mathrm{ee}^{-1} + \tau_\mathrm{eph}^{-1}$ with $\tau_\mathrm{ee}$ and $\tau_\mathrm{eph}$ given
by Eqs. (\ref{tauee}) and (\ref{tauGK}) respectively, and $T_H$ is the hot carrier temperature ($T_H \ll \Phi_B$) 
given in the units of energy by
\begin{equation}
 \label{lowTH}
 T_H=\hbar\bar\omega/ \ln\left[1  + \frac{D_0^2\bar\omega^3}{3\pi\rho v^4} \frac{1}{I_a \left(1- \tau_\mathrm{th} / \tau_\mathrm{cool}^\mathrm{nth}\right)}\right].
\end{equation}
Here, $D_0=11\, \mathrm{eV/\mathring{A}}$ is the deformation potential, $\rho=7.6\cdot 10^{-8}$ g/cm$^2$ is the mass density of graphene,
$\bar\omega\simeq 177$ meV is the average optical phonon frequency \cite{PRB2012low},
and $\tau_\mathrm{cool}^\mathrm{nth}$ is the nonthermalized photocarrier energy dissipation time.
The ratio $\tau_\mathrm{th} / \tau_\mathrm{cool}^\mathrm{nth}$ is shown in Fig. \ref{fig2} by red color,
but it can be calculated explicitly from Eqs. (\ref{tauee}), (\ref{tauGK}), and (\ref{taucool}).
Equation (\ref{Ratio}) is plotted in Fig. \ref{fig3} by red color demonstrating a relative decrease
of the thermionic component at lower excitation energies.

\section{Approximations involved}

In order to make our model analytically tractable we rely on the following assumptions:
(i) we assume intrinsic graphene, i.e. electron and hole concentrations are always the same;
(ii) both photocarrier thermalization and cooling are due to optical phonon emission; 
(iii) both electron ($T_H$) and lattice ($T_L$) temperatures are assumed to be much lower than optical phonon energy quanta $\hbar\omega_p$;
(iv) the latter is in turn assumed to be much lower than the excitation energy quantum $\hbar\omega$;
(v) we assume a linear response in absorbed radiation intensity $I_a$;
(vi) the radiation is assumed to be perfectly monochromatic so that electrons and holes are respectively
excited at energies $\varepsilon_\omega=\pm\hbar\omega/2$;
(vii) excitation energy is assumed to be low enough to allow for a conical band approximation with the linear dispersion $\varepsilon= \pm \hbar v k$ (see Fig. \ref{fig1}),
where $k$ is the electron wave vector.

Assumption (i) is well justified as long as photocarrier thermalization
is governed by optical phonon emission with high energy quanta ($\hbar\omega_p \sim 0.2$ eV) above the Fermi energy.
This is indeed the case in low-doped graphene where a small Fermi surface and momentum conservation severely constrain the phase space 
for acoustic phonon scattering so that higher order e-ph interaction terms must
be considered \cite{PRL2012song} to agree with measurements \cite{graham2013photocurrent,betz2013supercollision}.
The leading role of optical phonons in photocarrier thermalization
has been very recently confirmed by ab-initio calculations \cite{dinesh2018photocarrier}.
The assumptions (iii) and (iv) can be summarized in the inequality
 $T_{L,H} \ll \hbar\omega_p \ll \hbar\omega$.
Here, we have in mind that the optical excitation energy is of the order of 1 eV,
the optical phonon energy quanta are of the order of 100 meV in graphene, and
the lattice and photocarrier temperatures are both of the order of 10 meV. 
(The latter is of course somewhat higher than the former.)
The linear response assumption in $I_a$ suggests that the photocarrier thermalization and energy dissipation times
are both independent of the radiation intensity absorbed. 
The continuous-wave operation mode is consistent with the low radiation intensity
as well as with the narrow spectral width that allows for a $\delta$-shaped approximation.

The photocarrier emission across a 2D heterostructure shown in Fig. \ref{fig1}
strongly differs from conventional electron transport through bulk metal-semiconductor Schottky junctions \cite{SciRep2014zhang}.
The conventional Schottky junctions contain a so-called depletion region
--- a positively charged layer consisting of immobile ionized donors.
The depletion region extends over a micron inside the bulk semiconductor \cite{PRX2012tongay,di2016graphene}.
The depletion layer is mirrored by a very thin layer of opposite-sign charge at the metal surface
giving rise to a built-in electric field and a built-in potential.
In contrast to the Schottky contacts between graphene and a bulk semiconductor \cite{PRX2012tongay,trushin2018theory},
the 2D heterostructures considered here are so thin that the depletion layer just cannot fit inside the junction \cite{SciRep2014zhang}.
Moreover, 2D semiconductors contain much less impurities (hence, ionized donors) than their bulk parent crystals.
Thus, we assume that the built-in potential is negligible in true 2D heterostructures
so that the Schottky barrier height is determined solely by the 
the work function difference in graphene \cite{zhu2014heating} and a semiconductor \cite{ACSnano2016lee}  
modulated by an appropriate doping \cite{JPC2015chengjun}.

To simplify notation we focus on conduction band electrons ($\varepsilon >0$) and omit the band index.
The interband transitions are nevertheless allowed in the collision integrals.
The steady-state linear-response  nonequilibrium photoelectron distribution function $f_\varepsilon$
is a sum of a Fermi-Dirac distribution $f^{(0)}_\varepsilon$ (taken at temperature $T_H$) and a nonthermalized addition $f^{(1)}_\varepsilon$.
The distribution function must satisfy the following equation 
\begin{equation}
\label{th}
 G_\omega(\varepsilon) + {\cal I}_{ee}[f_\varepsilon] +  \sum_p {\cal I}_p[f_\varepsilon]=0,
\end{equation}
where ${\cal I}_{ee}[f_\varepsilon]$ is the e-e collision integral, 
${\cal I}_p[f_\varepsilon]$ is the e-ph collision integral for a phonon mode $p$,
and $G_\omega$ is the photoelectron generation rate.
The e-e collision integral for carriers in graphene has been analyzed in 
Refs. [\onlinecite{rana2007electron,PRBwinzer2012,JPCM2015song,PRB2013tomadin,Natcomm2013brida,PRB2016trushin}].
Here, we utilize a collinear-scattering approximation \cite{PRB2016trushin,PRB2008fritz} to keep the model analytically tractable.
The e-ph scattering has been investigated in Refs. [\onlinecite{PRL2009bistritzer,PRB2009kubakaddi,PRB2009tse,PRB2012low,PRL2012song,JPCM2015song}],
but here we employ our very recent analytical and numerical results from Ref. [\onlinecite{dinesh2018photocarrier}].
The photocarrier generation rate can be easily calculated using the golden-rule transition rate and reads \cite{PRB2015trushin}
\begin{equation}
 \label{G}
 G_\omega = I_a \frac{\pi v^2}{\omega^2}\delta\left(\varepsilon - \frac{\hbar\omega}{2} \right).
\end{equation}
Multiplying Eq. (\ref{G}) by energy and integrating it over occupied states one can see that
conduction band electrons absorb exactly one half of the intensity $I_a$, whereas another half is absorbed by valence band holes.

\section{Thermalization}

A general solution of Eq.~(\ref{th}) should contain the cascade terms generated 
when the optical phonons are emitted or absorbed \cite{PRB2011malic}.
We use the relaxation time approximation, i.e. we truncate the cascade solution to a single term
proportional to  $\delta\left(\varepsilon -  \hbar\omega/2\right)$.
Using this {\em Ansatz} along with the assumptions summarized above we have shown in Ref. [\onlinecite{PRB2016trushin}] that for intrinsic graphene
${\cal I}_{ee}[f_\varepsilon] = {\cal I}_{ee}[f_\varepsilon^{(1)}] \simeq - f_\varepsilon^{(1)}/\tau_\mathrm{ee}$,
where $\tau_\mathrm{ee}$ is given by
\begin{equation}
 \label{tauee}
\frac{1}{\tau_\mathrm{ee}} = \omega \frac{\tilde\alpha^2}{8\pi^3} \ln\left(\frac{1}{\tilde\alpha}\right).
\end{equation}
Here $\tilde\alpha=e^2/(\bar\epsilon \hbar v)$ is the effective fine-structure
constant for carriers in graphene determined by the effective dielectric constant $\bar\epsilon$.
For graphene on the most conventional substrates, $\tilde\alpha$ is typically 
between $0.3$ and $0.8$, hence, the photocarrier relaxation time scales with the photocarrier energy as 
$\tau_\mathrm{ee} \sim 1\,\mathrm{ps\cdot eV}/(\hbar\omega/2)$,
which is in perfect agreement with the numerical results (see Fig. 6 in Ref. [\onlinecite{Natcomm2016mihnev}]).
Using the same assumptions we write the e-ph collision integral as a sum of two terms \cite{dinesh2018photocarrier}, i.e.,
${\cal I}_p[f_\varepsilon]={\cal I}_p[f^{(0)}_\varepsilon]+{\cal I}_p[f^{(0)}_\varepsilon,f^{(1)}_\varepsilon]+o^2(I_a)$, where
\begin{eqnarray}
 \nonumber  &&  {\cal I}_p[f^{(0)}_\varepsilon] =  \frac{W_p}{\hbar^3v^2}\left[
 |\varepsilon + \hbar\omega_p| f^{(0)}_{\varepsilon+\hbar\omega_p} f^{(0)}_{-\varepsilon} \right. \\
  \label{Ieph3} && \left. - |\varepsilon - \hbar\omega_p|f^{(0)}_{-\varepsilon+\hbar\omega_p} f^{(0)}_\varepsilon  \right],\\
 \nonumber && {\cal I}_p[f^{(0)}_\varepsilon,f^{(1)}_\varepsilon] = \\
 \nonumber && \frac{W_p}{\hbar^3v^2}\left[
 |\varepsilon + \hbar\omega_p| \left(f^{(1)}_{\varepsilon+\hbar\omega_p} f^{(0)}_{-\varepsilon} - f^{(1)}_\varepsilon f^{(0)}_{\varepsilon+\hbar\omega_p} \right) \right. \\
 \label{Ieph4}  && -  \left. |\varepsilon - \hbar\omega_p| \left(f^{(1)}_\varepsilon f^{(0)}_{-\varepsilon+\hbar\omega_p} 
 - f^{(1)}_{\varepsilon-\hbar\omega_p} f^{(0)}_\varepsilon \right) \right].
\end{eqnarray}
Here $W_p$ is the golden-rule e-ph interaction given by $W_p= (\hbar D_p^2 F_p)/(2\rho\omega_p)$,
where $D_p$ is the deformation potential, and $F_p$ is a dimensionless geometric factor 
($F_\Gamma=1$ for the $\Gamma$-mode and $F_K=1/2$ for the $K$-mode) \cite{PRB2012low}.
Note that $W_p$ does not depend on electron momentum in our model.
Equation (\ref{Ieph3}) does not contain the nonthermalized term $f^{(1)}_\varepsilon$, and, therefore, does not contribute to thermalization. 
Equation (\ref{Ieph3}) is however responsible for cooling of thermalized carriers, see the next section.
Equation (\ref{Ieph4}) contributes to both thermalization and energy dissipation from nonthermalized carriers.
This contrasts to the e-e collision integral, where elastic e-e scattering does not dissipate photocarrier energy.
Using Ref. [\onlinecite{dinesh2018photocarrier}], we rewrite Eq. (\ref{Ieph4}) as
${\cal I}_p[f^{(0)}_\varepsilon,f^{(1)}_\varepsilon] \simeq - f_\varepsilon^{(1)}/\tau_\mathrm{eph}$, where 
\begin{eqnarray}
&& \nonumber
\frac{1}{\tau_\mathrm{eph}}=\sum\limits_p \frac{W_p}{\hbar^3v^2} \left(\frac{\hbar\omega}{2} - \hbar\omega_p  \right)  \\
  && = \frac{D_0^2 }{\rho\hbar v^2}\left(\frac{\omega}{4\omega_0} -1 \right).
    \label{tauGK} 
 \end{eqnarray}
Here we take into account the two most important phonon modes \cite{PRB2012low}, $p=\Gamma,K$,  denote 
$ 1/\omega_0=1/\omega_\Gamma+1/\omega_K$,
$D_\Gamma=D_0$,  $D_K=\sqrt{2}D_0$, and make use of the explicit values for $F_{\Gamma,K}$, see Appendix B in Ref. [\onlinecite{PRB2012low}].
We estimate $\tau_\mathrm{eph}$ to be a few tens of fs at excitation energies within the visible spectrum, i.e.,
$\tau_\mathrm{eph}\ll \tau_\mathrm{ee}$ so that the thermalization rate $\tau_\mathrm{th}^{-1} = \tau_\mathrm{ee}^{-1} + \tau_\mathrm{eph}^{-1}$ 
is mostly determined by phonon emission.
This agrees with the recent ultrafast pump-probe spectroscopy measurements \cite{PRB2015trushin},
where the photocarriers excited at $\hbar\omega=1.62$ eV turn out to be thermalized within 60 fs due to e-ph scattering.
At lower excitation energies, $\tau_\mathrm{eph}$ increases and thermalization turns out to be remarkably slow.
Once $\omega$ approaches $\omega_0$ the thermalization occurs at much longer (ps) time scale. 
This prediction has also been confirmed recently by ab-initio calculations \cite{dinesh2018photocarrier} and 
by means of pump-probe spectroscopy near the optical phonon emission bottleneck \cite{PRL2016otto}.
Hence the outcomes of our explicitly solvable thermalization model are consistent
with the measurements \cite{PRB2015trushin,PRL2016otto} as well as with the numerical calculations \cite{Natcomm2016mihnev,dinesh2018photocarrier}.
The resulting nonthermalized distribution function follows from Eqs. (\ref{th}--\ref{tauee}, \ref{tauGK}), and can be written as
\begin{equation}
 \label{solution}
 f^{(1)}_\varepsilon = \tau_\mathrm{th} I_a \frac{\pi v^2}{\omega^2} \delta\left(\varepsilon  -  \hbar\omega/2\right).
\end{equation}

\section{Cooling}

The hot Fermi-Dirac distribution function contains the electron temperature $T_H$ which is still to be determined.
To do that we first deduce the electron energy balance equation by multiplying Eq. (\ref{th}) with energy
and integrating it in momentum space. We know that e-e collisions do not dissipate energy and hence do not contribute to this equation.
In our model, the hot electron energy is dissipated by means of optical phonon emission only. 
Hence, the energy dissipation rate is determined by balancing the radiation intensity absorbed by electrons from an external source
with the intensity dissipated from electrons by means of optical phonons emission.
Most importantly, {\em both} thermalized and nonthermalized electrons can emit phonons.
Hence, the energy balance equation contains {\em two} dissipating terms representing the phonon energy emitted by thermalized and nonthermalized electrons.
For a given spin/valley channel this equation can be written as
\begin{equation}
\label{bal-start}
I_a/8 + I_{T_H} + I_\omega =0.
\end{equation}
Here
\begin{equation}
I_\omega = \sum\limits_p \int\frac{d^2k}{(2\pi)^2} \varepsilon {\cal I}_p[f^{(0)}_\varepsilon,f^{(1)}_\varepsilon],
 \label{Iomega}
\end{equation}
\begin{equation}
I_{T_H} = \sum\limits_p \int\frac{d^2k}{(2\pi)^2} \varepsilon {\cal I}_p[f^{(0)}_\varepsilon],
\end{equation}
where ${\cal I}_p[f^{(0)}_\varepsilon]$ and ${\cal I}_p[f^{(0)}_\varepsilon,f^{(1)}_\varepsilon]$ are given by Eqs.~(\ref{Ieph3})
and (\ref{Ieph4}), respectively. 
Note that each term in $I_\omega$  contains $f^{(1)}_\varepsilon$ given by (\ref{solution}) so that
we make use of the $\delta$-function and calculate the integral over $\varepsilon$ explicitly.
Using the assumptions (iii) and (iv) we can write $I_\omega$ simply as
$I_\omega = -\tau_\mathrm{th} I_a/8\tau_\mathrm{cool}^\mathrm{nth}$, where
\begin{eqnarray}
 \nonumber \frac{1}{\tau_\mathrm{cool}^\mathrm{nth}} & = & \sum\limits_p \frac{W_p \omega_p}{\hbar^2 v^2}\left(1- \frac{2\omega_p}{\omega}\right) \\
 &=& \frac{D_0^2 }{\rho\hbar v^2}\left(1 - \frac{\omega_\Gamma + \omega_K}{\omega} \right).
 \label{taucool}
\end{eqnarray}
Here, similar to Eq. (\ref{tauGK}), only $p=\Gamma, K$ phonon modes are taken into account.
The index ``$\mathrm{nth}$'' stands for ``nonthermalized'', i.e., Eq.~(\ref{taucool}) represents the energy dissipation rate for nonthermalized 
photoexcited electrons. For thermalized electrons we can write
$I_{T_H}=D_0^2\left[\omega_\Gamma^3 F(\beta_\Gamma) + \omega_K^3 F(\beta_K)\right]/(4\pi\rho v^4)$,
where
\begin{eqnarray}
 \label{Fp}
 \nonumber && F(\beta_p)=\int\limits_0^\infty dx x^2\left[\frac{x+1}{\left(1+ \mathrm{e}^{(x+1)\beta_p} \right)\left(1+ \mathrm{e}^{-x\beta_p} \right) } \right. \\
 && \left. -\frac{|x-1|}{\left(1+ \mathrm{e}^{(1-x)\beta_p} \right)\left(1+ \mathrm{e}^{x\beta_p} \right) } \right],
\end{eqnarray}
and $\beta_p=\hbar\omega_p/T_H$. 
In the high-$T_H$ regime ($T_H\gg \hbar\omega_p$) we have 
\begin{equation}
 \label{highTHFp}
 F(\beta_p)\simeq - \frac{\pi^2}{6} \frac{1}{\beta_p^3},
\end{equation}
hence, the energy balance equation reads
\begin{equation}
 \label{highTbalance}
 \frac{I_a}{8} \left(1-\frac{\tau_\mathrm{th}}{\tau_\mathrm{cool}^\mathrm{nth}}\right) -
 \frac{\pi D_0^2 T_H^3}{12\hbar^3 \rho v^4} =0,
 \end{equation}
and the hot electron temperature is given by
\begin{equation}
 \label{highTH}
 T_H=\hbar\omega_H,\quad \omega_H= \left[\frac{3\rho v^4}{2\pi D_0^2}I_a\left(1- \frac{\tau_\mathrm{th}}{\tau_\mathrm{cool}^\mathrm{nth}}\right) \right]^{\frac{1}{3}}.
\end{equation}
In the low-$T_H$ limit ($\beta_p\gg 1$) we have 
\begin{equation}
 \label{lowTHFp}
 F(\beta_p)\simeq - \frac{1}{12} \frac{1}{\mathrm{e}^{\beta_p} - 1}.
\end{equation}
The energy balance equation is not explicitly solvable in this limit, however, assuming some average optical phonon frequency $\bar\omega$  
(i.e., we set $\omega_K=\omega_\Gamma=\bar\omega$), we obtain Eq.~(\ref{lowTH}).
Hence, we have now an explicit expression for the steady-state photoelectron distribution function in intrinsic graphene:
$f_\varepsilon=f^{(0)}_\varepsilon(T_H)+f^{(1)}_\varepsilon(\tau_\mathrm{th})$,
where $T_H$ is given by Eq.~(\ref{lowTH}), and $f^{(1)}_\varepsilon(\tau_\mathrm{th})$ is given by Eq.~(\ref{solution}).

\section{Transport}

We now calculate the electron current density across the junction by setting the bias voltage to zero and 
assuming an elastic but momentum nonconserving interlayer transport \cite{NanoLett2015rodriguez}.
The electron hopping probability is calculated using a $\delta$-shaped disorder potential
$\epsilon_d r_d^2\delta(\mathbf{r})$ written in terms of the on-site energy \cite{PRL2010wehling} $\epsilon_d=0.26$ eV and radius $r_d=1.7\, \mathring{\mathrm{A}}$,
which is about one half of the interlayer distance \cite{Nanoscale2011ma,JPC2015chengjun}.
The short-range disorder concentration, $n_d\simeq 10^{11}$ cm$^{-2}$, is deduced from the carrier mobility, $\mu\simeq 10^4$ cm$^2$/(Vs), 
using the resonant scattering model \cite{PRL2010wehling,PRB2014sopik}.
The golden-rule interlayer transmission probability due to a single scatterer then reads
\begin{eqnarray}
 w_{\mathbf{q}\mathbf{k}}= \frac{2\pi}{\hbar}\frac{\epsilon_d^2 r_d^4}{ L^4} \delta\left(\frac{\hbar^2 q^2}{2m^*} - \hbar v k\right),
\end{eqnarray}
where $\mathbf{q}$ ($\mathbf{k}$) is the in-plane electron momentum in graphene (semiconductor), 
$m^*/m_0\simeq 0.3$ is the ratio between effective ($m^*$) and free ($m_0$) electron masses on the semiconductor side \cite{PRB2012cheiwchanchamnangij},
and $L$ is the junction size. (The latter appears due to the wave function normalization.)
The current density from graphene (G) to a semiconductor (S) can be written as
$J_{G\to S } = e g_{s\nu} n_d \sum_{\mathbf{q},\mathbf{k}} w_{\mathbf{q}\mathbf{k}} f_\mathbf{k} (1-f_\mathbf{q})$,
whereas in the reversed direction it reads
$J_{S \to G} = e g_{s\nu} n_d \sum_{\mathbf{q},\mathbf{k}} w_{\mathbf{q}\mathbf{k}} f_\mathbf{q} (1-f_\mathbf{k})$.
Here $f_\mathbf{k}$ ($f_\mathbf{q}$) are the carrier distribution functions in graphene (semiconductor), and $g_{s\nu}$ is the spin/valley degeneracy
($g_{s\nu}=4$ for graphene/MoS$_2$ junctions).
If the electron occupation on the semiconductor side is much smaller than in graphene ($f_\mathbf{k} \gg f_\mathbf{q}$), then the total current density
$J=J_{G\to S }  - J_{S\to G }$ can be estimated as
\begin{equation}
\label{c1}
J = \frac{2\pi e}{\hbar}g_{s\nu}n_d\epsilon_d^2 r_d^4 \int\limits_0^\infty d\varepsilon D_S\left(\varepsilon\right) D_G\left(\varepsilon\right)
\left(  f^{(0)}_\varepsilon +  f^{(1)}_\varepsilon \right),
\end{equation}
where $D_G =\varepsilon/(2\pi\hbar^2 v^2)$ and $D_S=m^*\theta(\varepsilon-\Phi_B)/(2\pi \hbar^2)$
are the densities of states for electrons in graphene and 2D semiconductor, respectively, with $\theta(\varepsilon-\Phi_B)$ being the Heaviside steplike function.
If $\Phi_B \gg T_H$ but $\Phi_B < \hbar\omega/2$, then the integral is easy to take, and the total photoresponsivity $R_\mathrm{tot}=J/I_a$ can be written as
\begin{equation}
\label{photores}
R_\mathrm{tot} = \frac{e}{\hbar\omega}\frac{\tau_\mathrm{th}}{\tau_d}  + 
\frac{2e}{\pi}\frac{\Phi_B T_H}{\hbar^2 v^2 \tau_d I_a}\mathrm{e}^{-\frac{\Phi_B}{T_H}},
\end{equation}
where
\begin{equation}
 \label{hoprate}
 \frac{1}{\tau_d}=\frac{m^* n_d \epsilon_d^2 r_d^4}{\hbar^3}
\end{equation}
is the interlayer hopping rate, and $\tau_d \sim 1 $ ns for the parameter values provided above.
The photoresponsivity formally increases with disorder strength because the scattering potential is taken into account perturbatively in our model
that makes Eq. (\ref{photores}) inapplicable in the limit of $\tau_d\to 0$.
The ratio $\tau_\mathrm{th}/\tau_d \lesssim 10^{-3}$, i.e., 
the interlayer hopping is much slower than thermalization that confirms the golden-rule applicability to the interlayer transport description.

The first and second terms in Eq. (\ref{photores}) represent, respectively, the photoelectric 
($R_\mathrm{ph}$) and thermionic ($R_\mathrm{th}$) contributions to the photoresponsivity, and their ratio is given by Eq. (\ref{Ratio}).
Using the Schottky barrier height of about $0.3$ eV, 
and absorbed intensity of a few W/mm$^2$ [\onlinecite{NanoLett2016goykhman},\onlinecite{NanoLett2015rodriguez}]
we plot the total photoresponsivity in Fig. \ref{fig3} to show that it substantially increases at longer excitation wavelengths thanks
to higher photocarrier occupation contributing to the photocurrent near the excitation energy $\hbar\omega/2$.

\section{Discussion}

We find that the steady-state photocarrier occupation 
generated by $I_a$ of the order of $10$ W/mm$^2$ within the optical spectral range can be well approximated just by the hot Fermi-Dirac distribution
$f^{(0)}_\varepsilon$ because $\tau_\mathrm{th}$ is only a few tens of fs, and nearly all the photoexcited carriers are thermalized (see Fig. \ref{fig2}).
Hence, the photoresponsivity is dominated by  thermionic emission, i.e., by the second term in Eq. (\ref{photores}).
This is not the case when the photocarriers are excited within the near infrared region 
because $\tau_\mathrm{th}$ gets longer and, as a consequence, a lager fraction of photocarriers remains in a nonthermalized steady state (see Fig. \ref{fig2}).
At the same time $T_H$ is getting lower because $\tau_\mathrm{th}$ and $\tau_\mathrm{cool}^\mathrm{nth}$
turn out to be of the same order of magnitude (a few hundreds of fs). Hence, the photoresponsivity is governed by the first term in Eq. (\ref{photores})
at near infrared excitation energies and below (see Fig. \ref{fig3}).
At  $I_a \lesssim 1$ W/mm$^2$ the photoelectric emission dominates within the whole excitation region shown in Fig. \ref{fig3}
because the steady-state carrier temperature $T_H$ is too low for an efficient thermionic regime.

To justify the absorbed intensity of a few W/mm$^2$ we assume the guided mode approach \cite{koppens2014photodetectors}
that raises the optical absorption in graphene well beyond 2.3\% and, by increasing the interaction length, 100\% light power can be absorbed \cite{pospischil2013cmos}.
The absorbed energy is redistributed between non-thermalized and thermalized carriers according to the ratio
$(\tau_\mathrm{th}/\tau_\mathrm{cool}^\mathrm{nth})\div(1-\tau_\mathrm{th}/\tau_\mathrm{cool}^\mathrm{nth})$
that can be estimated as $1\div 9$ for $\hbar\omega \sim 3.5$ eV
and $1\div 1$ for $\hbar\omega \sim 0.8$ eV. The energy fraction transfered to a semiconductor
is given by the ratio $\tau_\mathrm{th}/\tau_d$ that is less than $0.1\%$ of the absorbed intensity
and is, therefore, neglected in the intensity balance equation (\ref{bal-start}).
This is also the reason why the responsivity predicted here is lower than measured in the state of the art photodetectors based 
on graphene-semiconductor junctions \cite{NatPhot2013gan,NanoLett2016goykhman,ACSNano2016defazio}.
We consider photoresponsivity of an ideal 2D junction at zero external voltage where the only
mechanism causing interlayer electron hopping is due to weak scattering on the short-range disorder.
A small bias could facilitate the interlayer transport increasing the photoresponsivity
but it would not compromise photocarrier thermalization dynamics we have considered.

\section{Outlook}

The model predictions outlined above require experimental verification. 
First, photoresponsivity should increase with excitation wave length.
This prediction can easily be tested as graphene's optical absorption does not depend on excitation energy, hence,
the dependence of photoresponsivity on radiation wavelength is solely determined by photocurrent. 
One should however make sure that the photocarriers are always excited above the Schottky barrier and below the semiconductor band gap.
Second, graphene samples with lower mobilities should demonstrate a somewhat higher electron hopping rate into the semiconductor
producing higher photocurrent and resulting in higher photoresponsivity.
To verify the model one would need a set of graphene samples with various in-plane mobilities.
Needless to say, the disorder should not be too high to block the electron transport completely.

We envisage possible applications of this model in designing future graphene-based optoelectronic and photovoltaic devices
where a quick assessment of photocarrier steady-state temperature and energy losses is needed \cite{frontiers2014,C3EE42098A}.
The most obvious application could be photodetection at the telecommunication wavelength ($1.55\mathrm{\mu m}$) determined
by the radiation absorption characteristics of the fiber \cite{keiserbook}.
While conventional semiconductor-based photodetectors can work in the visible spectral range they are not suitable for detecting 
at the telecommunication wavelength because the corresponding photon energy of 0.8 eV is below the typical semiconductor band gap size and, therefore,
far too low to gain a substantial photoresponse. Graphene-semiconductor junctions \cite{SciRep2014zhang,ryzhii2017nonlinear,roy2013graphene} can potentially solve this problem:
The photocarriers can be excited in graphene and effectively transferred to the semiconductor side 
by means of the photoelectric effect due to the longer thermalization time for near infrared excitations. 

\acknowledgments
This work has been supported by the Director's Senior Research Fellowship 
from the Centre for Advanced 2D Materials at the National University of Singapore
(National Research Foundation Singapore Medium Sized Centre Programme R-723-000-001-281).
Multiple discussions with Shi-Jun Liang are acknowledged.

\bibliography{thermalization.bib}

\end{document}